# METRIC-SPACE ANALYSIS OF SPIKE TRAINS: THEORY, ALGORITHMS, AND APPLICATION

Jonathan D. Victor and Keith P. Purpura

Department of Neurology and Neuroscience
Cornell University Medical College
1300 York Avenue
New York, NY 10021

Address correspondence to:

Jonathan D. Victor
Department of Neurology and Neuroscience
Cornell University Medical College
1300 York Avenue
New York, New York 10021

voice: (212) 746-2343
fax: (212) 746-8984
email: jdvicto@med.cornell.edu

ABSTRACT

We present the mathematical basis of a new approach to the analysis of temporal coding. The foundation of the approach is the construction of several families of novel distances (metrics) between neuronal impulse trains. In contrast to most previous approaches to the analysis of temporal coding, the present approach does not attempt to embed impulse trains in a vector space, and does not assume a Euclidean notion of distance. Rather, the proposed metrics formalize physiologically-based hypotheses for what aspects of the firing pattern might be stimulus-dependent, and make essential use of the point process nature of neural discharges. We show that these families of metrics endow the space of impulse trains with related but inequivalent topological structures. We show how these metrics can be used to determine whether a set of observed responses have stimulus-dependent temporal structure without a vector-space embedding. We show how multidimensional scaling can be used to assess the similarity of these metrics to Euclidean distances. For two of these families of metrics (one based on spike times and one based on spike intervals), we present highly efficient computational algorithms for calculating the distances. We illustrate these ideas by application to artificial datasets and to recordings from auditory and visual cortex.

## INTRODUCTION

Recent neurophysiologic studies in vision (12, 25, 30, 33, 35, 47), audition (1, 13, 28), and olfaction (22, 41) have provided convincing evidence that sensory information is represented by the temporal structure of a neural discharge, as well as by the number of spikes in the response. Although a neuronal discharge is fundamentally a point process, many approaches to the analysis of temporal structure rely on binning the spike trains and on adopting methods appropriate for multivariate data (10) or continuous signals. There are two potential drawbacks to such approaches (47). One problem is that an adequate time resolution over a reasonable analysis interval requires an embedding in a high-dimensional vector space which is only sparsely sampled by the dataset. More fundamentally, this approach is less than optimal because the vector space approach (i) treats all coordinates on an equal footing, and thus ignores the sequential nature of time; and (ii) assumes that the space of spike trains has a Euclidean geometry. One might argue that an assumption of an underlying Euclidean geometry is justified in instances in which there is an approximately Euclidean "perceptual space" -- such as representations of color -- but it is difficult to justify this assumption when the geometry of the perceptual space is unknown -- e.g., a representation of "objects".

Methods which do not require vector-space embeddings have been used to examine temporal coding, but currently-available methods have other drawbacks. A neural network scheme (19, 28) for the classification of spike discharges can surmount the temporal resolution problem. However, inferences concerning the nature of the temporal code are not straightforward, since they require an understanding both of the manner in which spike trains are represented (19) and the parameters of the neural network. Other approaches deal explicitly with spike trains as point processes, but these methods focus on correlations among discharges (31, 32), the pattern of interspike intervals (34), or the identification of similar segments of spike discharges (3), rather than on a global analysis of how the pattern of the discharge depends on the stimulus.

We recently (47) used a novel approach to investigate temporal coding in the primate visual cortex. One major distinction between the present approach and many previous approaches is that it provides a global analysis of how the discharge pattern depends on the stimulus, without the need for an embedding in a vector space, or an assumption of a Euclidean (or near-Euclidean) geometry for the set of spike trains. The philosophy behind this approach is to exploit what is known about the biological significance of the temporal pattern of nerve impulses to construct a specifically appropriate mathematical structure, rather than to adapt general-purpose methods of signal processing. The purpose of this paper is to describe the mathematical basis of this approach in detail.

# RESULTS

## *Overview*

For the reasons described in the Introduction, we construct a method to analyze the temporal structure of spike trains based only on the bare essentials: an abstract set of points (the spike trains) and a self-consistent definition of distances between pairs of these points. In formal mathematical terms, we consider the spike trains and the notion of distance to define a metric space (11), a topological structure substantially more general than a vector space with a Euclidean distance. The extent to which our construct indeed corresponds to a Euclidean distance in a vector space will be determined empirically, rather than assumed.

We will consider several families of metrics. Each metric determines a candidate geometry for the space of spike trains. Stimulus-dependent clustering will be assessed relative to each candidate geometry, without recourse to further mathematical structure.

The metrics we consider are based on intuitions concerning possible biological underpinnings of temporal coding. The first family of metrics, denoted $D^{spike}[q]$, emphasizes the significance of the absolute timing of individual impulses. The rationale for this family of metrics is that under some circumstances, a cortical neuron may behave like a coincidence detector (2, 27, 42, 43), but the effective resolution of this coincidence detector is uncertain. Within the resolution of this coincidence detector, the effect of a spike train on other cortical neurons will depend on the absolute timing of its impulses, rather than on the number of spikes within a given interval.

A second family of metrics, denoted $D^{interval}[q]$, emphasizes the duration of interspike intervals. The rationale in this case is that the effect of an action potential can depend critically on the length of the time since earlier action potentials (39). Possible biological substrates for this dependence include the NMDA receptor and $Ca^{2+}$ channels whose behavior is sensitive to the pattern of interspike intervals (6, 21, 36). While it is trivially true that the absolute spike times determine the interspike intervals (and *vice-versa*, with the notion that the first "interspike interval" is the interval between the onset of data collection and the first spike), it is not true that the distance between two spike trains, in the sense of $D^{spike}$, determines the distance between the trains in the sense of $D^{interval}$.

Finally, we will consider a third family of metrics, $D^{motif}$, which is motivated by the notion that a "motif," or a temporal pattern of a subset of spikes (3), may be of physiological significance. This family of metrics is a natural formal extension of $D^{spike}$ and $D^{interval}$. However, implementation of our analysis for $D^{motif}$ is hindered by the lack of availability of efficient algorithms.

Within each of these families, the specific metrics that we will consider ($D^{spike}[q]$, $D^{interval}[q]$, and $D^{motif}[q]$) depend on a parameter $q$, which expresses the sensitivity of the metric to temporal pattern. The parameter $q$ has units of ($sec^{-1}$) and represents the relative cost to "move" a spike (for $D^{spike}[q]$), to change the duration of an interval (for $D^{interval}[q]$), or to translate a motif (for $D^{motif}[q]$), compared to the cost of inserting or deleting a spike. For $q = 0$, each of these metrics reduce to a distance based solely on counting spikes. Thus, temporal coding will manifest itself as more reliable clustering for some values of $q > 0$ than for $q = 0$. For sufficiently large values of $q$, we anticipate a decrease in systematic clustering, since the infinitesimally precise timing of impulses or intervals cannot possibly carry biological information. Thus, our method provides two characterizations of temporal coding: the amount of systematic clustering seen with $q > 0$ will indicate the extent to which absolute spike times ($D^{spike}[q]$), spike intervals ($D^{interval}[q]$), or subsets of spikes ($D^{motif}[q]$) depend on the stimulus, and the value of $q$ for which the clustering is greatest. The latter characterization will indicate the temporal resolution of the coding.

*Mathematical framework and definition of metrics*

A metric space (11) is a set of points (here, spike trains, to be denoted $S_a$, $S_b$, ...) along with a metric $D$, which is a mapping from *pairs* of spike trains to the real numbers. In order for $D$ to be a metric, it must (i) always be positive except for the trivial case $D(S, S) = 0$, (ii) be a symmetric function $[D(S_a, S_b) = D(S_b, S_a)]$, and (iii) satisfy the triangle inequality

$$D(S_a, S_c) \leq D(S_a, S_b) + D(S_b, S_c) \quad . \tag{1}$$

With these conditions satisfied, the function $D$ can be thought of as specifying a distance.

A spike train $S$ is a sequence of times $t_1, t_2, ..., t_k$, with $0 \leq t_1 < t_2 < ... < t_k$ , and will be denoted $S = \{t_1, t_2, ..., t_k\}$. We will define a metric $D(S_a, S_b)$ as the minimum "cost" required to transform the spike train $S_a$ into the spike train $S_b$ via a path of elementary steps. The cost assigned to a path of steps is the sum of the costs assigned to each of the elementary steps. Formally,

$$D(S_a, S_b) = \operatorname*{glb}_{S_0, S_1, ..., S_r} \sum K(S_j, S_{j-1}) \;, \quad where \; S_0 = S_a, \; S_r = S_b, \tag{2}$$

and where $K(S_j, S_{j-1})$ is equal to the cost of an elementary step from $S_j$ to $S_{j-1}$. $K(S_j, S_{j-1})$ is required to be non-negative and symmetric. $S_1, S_2, ..., S_{r-1}$ represent intermediate spike trains along the path from $S_a$ ( = $S_0$) to $S_b$ ( = $S_r$). There is no need to specify that there is a path which *achieves* the minimum total cost, and thus we use the notation "glb" (= greatest lower bound) rather than "min" in eq. (2)).

Generally, functions of the form specified by eq. (2) will satisfy the above three conditions (i) - (iii), and thus qualify as metrics. The symmetry of $D$ is inherited from the postulated symmetry of the cost function $K$. The triangle inequality (1) is automatically satisfied, because one path from $S_a$ to $S_c$ is the cost-minimizing path from $S_a$ to $S_b$, followed by the cost-minimizing path from $S_b$ to $S_c$. However, depending on the choice of the cost function $K$, there may be distinct spike trains $S_a$ and $S_b$ for which $D(S_a, S_b) = 0$, and thus condition (i) would not be satisfied. This may be remedied (11) by considering the space of "equivalence classes of spike trains," where the equivalence class which includes $S$ is the set of spike trains whose distance from $S$ is zero. The function defined by eq. (2) now becomes a valid metric on equivalence classes of spike trains. The space of equivalence classes of spike trains is always a metric space, and a specification of the allowed elementary steps and their associated costs always provides a metric.

The nature of the metric defined by eq. (2) is determined by the allowed elementary steps and the costs that are assigned to them. For all of the metrics we consider here, the allowed elementary steps will always include adding a single spike or deleting a single spike. These steps will be assigned a cost of 1. This serves to ensure that there exists at least one path between any two spike trains. However, a metric which has only these allowed steps will see all spikes as equally different from each other, unless they occur at precisely the same time. To use the distances defined by eq. (2) to express more biologically-plausible notions of distance, additional kinds of elementary steps must be introduced.

A metric based on spike times. We first create a family of metrics whose distances reflect similar times of occurrence of impulses. This family, which we denote by $D^{spike}[q]$, has one kind of step in addition to spike insertion and deletion. This second kind of step is based on concatenation of a continuum of infinitesimal steps, each of which shift a single spike in time by an infinitesimal amount $dt$. The cost associated with this infinitesimal step is asserted to be $qdt$, where $q$ is a parameter with units $sec^{-1}$. Combining a continuum of these steps (each operating on the same spike) leads to a shift of a spike by an amount $\Delta t$, with the associated cost $q|\Delta t|$.

One extreme instance of this metric occurs if the cost/sec $q$ is set to zero. In this case, elementary steps which shift the position in time of a spike are free, and costs are associated only with adding or deleting spikes. It follows that the distance between these two spike trains in $D^{spike}[q]$ is the difference in the number of spikes, which we denote by the "spike count" metric $D^{count}$.

To gain some insight into $D^{spike}[q]$ for $q > 0$, consider two spike trains $\{a\}$ and $\{b\}$, each of which consist of only a single spike. There are two paths to consider in applying eq. (2). The path which consists of moving the solitary spike has a cost of $q|a - b|$. The path which consists of deleting the spike from $S_a$ and then inserting a spike to form $S_b$, and has cost of 2. It is cheaper to delete the spike and reinsert it than it is to move it, provided that $|a - b| > 2/q$. Thus, in the limit that $q$ is very large, the distance between two spike trains $S_a = \{a_1, a_2, ..., a_m\}$ and $S_b = \{b_1, b_2, ..., b_n\}$ is $m + n - 2c$, where $c$ is the number of spike times in $S_a \cap S_b$. In essence, $D^{count} = D^{spike}[0]$ ignores the time of occurrence of the spikes, while $D^{spike}[\infty]$ considers any difference in time of occurrence to constitute a "different" spike. For the metric $D^{spike}[q]$, displacing a spike by a distance $1/q$ is equal in cost to deleting it altogether, and displacing a spike by a smaller distance results in a spike train which is similar, but not identical. That is, $1/q$ is a measure of the temporal resolution of the metric. Equivalently, one can consider $q$ to be a measure of the precision of the temporal coding.

A metric based on interspike intervals. We now consider a second family of metrics, $D^{interval}[q]$, which depends on interspike intervals in much the same way that $D^{spike}[q]$ depends on spike times. For $D^{interval}[q]$, the second kind of elementary step is a concatenation of a continuum of infinitesimal steps, each of which consists of changing the length *of an interspike interval* by an infinitesimal amount $dt$. This step has cost $qdt$. A change in the length of an interspike interval necessarily changes the time of occurrence of all subsequent spikes. This is in contrast to the elementary step of $D^{spike}[q]$, in which only one spike time is changed, but two intervals are modified (the intervals immediately preceding and following the shifted spike).

In the two limiting cases of $q = 0$ and $q = \infty$, the metric $D^{interval}[q]$ is essentially the same as the metric $D^{spike}[q]$, because both depend only on the number of spikes. However, for intermediate values of $q$, the two metrics can have quite different behavior. This is because $D^{interval}[q]$ is sensitive to the pattern of interspike intervals, while $D^{spike}[q]$ is sensitive to absolute spike times. Consequently, $D^{interval}[q]$ can distinguish firing patterns that $D^{spike}[q]$ cannot. For example (as we will see below), $D^{interval}[q]$ can distinguish a pattern of intervals with a chaotic nonlinear recursion (34) from a renewal process with equal interval statistics; $D^{spike}[q]$

cannot make this distinction.

A technical detail, which concerns the initial and final intervals, arises in implementing the metric $D^{interval}[q]$. A spike train $S = \{t_1, t_2, ..., t_k\}$ on a segment $[0, T]$ unambiguously defines the $k$-1 interior intervals $t_2 - t_1, ..., t_k - t_{k-1}$. However, the initial and final intervals are of uncertain length: the initial interval is at least of length $t_1$, but may be longer (since the spike immediately preceding the spike at $t_1$ was not recorded). Similarly, the final interval is at least of length $T - t_k$, but may be longer. There are several ways to proceed to define a well-defined metric, each of which could be considered to be a variant of $D^{interval}[q]$. For example, the initial and final indeterminate intervals may simply be ignored -- creating a metric which we designate as $D^{interval:ign}[q]$. Alternatively, one may place an auxiliary leading spike in both trains at time 0 and a second auxiliary trailing spike in both trains at time $T$ -- creating a metric which we designate as $D^{interval:fix}[q]$. A third alternative is that the times of the auxiliary leading spikes inserted into the two trains can be allowed to vary independently (in the interval $[-\infty, 0]$ for the leading spike and the interval $[T, \infty]$ for the trailing spike), to minimize the distance of eq. (2) -- creating a metric which we designate as $D^{interval:min}[q]$. In general, these variations have only a minimal effect on the analysis of temporal structure, as would be expected since they are essentially end-effects. In the calculations presented in this paper, we used $D^{interval:fix}[q]$ (an auxiliary spikes inserted at time 0 and time $T$) and $D^{interval:min}[q]$ (auxiliary leading and trailing spikes were inserted in both trains at positions which minimized the distance of eq. (2)).

A metric based on subsets of spikes. The third family of metrics, $D^{motif}[q]$ is motivated by the notion that a "motif," or a temporal pattern of a subset of spikes (3), may be of physiological significance. This metric is constructed as a generalization of $D^{spike}[q]$. As in $D^{spike}[q]$, the first kind of step consists of adding a single spike, or deleting a single spike, and has a cost of 1. The second kind of step is again based on concatenation of a continuum of infinitesimal steps, but the infinitesimal steps allow joint shifting of any number of spikes, all in the same direction by the same amount *dt*. That is, the cost of a step from a spike train $S = \{t_1, t_2, ..., t_k\}$ to a spike train $S' = \{t'_1, t'_2, ..., t'_k\}$, where each $t'_j$ is either $t_j$ or $t_j + \Delta t$, has the cost $q|\Delta t|$. This metric is more closely related to $D^{interval}[q]$ than to $D^{spike}[q]$, in that an elementary step which shifts a contiguous subset of $N$ spike times changes only two intervals (the end intervals) and would thus have cost $2q|\Delta t|$ in $D^{interval}[q]$, but would have cost $Nq|\Delta t|$ in $D^{spike}[q]$. However, the metric $D^{motif}[q]$ is distinct from $D^{interval}[2q]$, in that $D^{motif}[q]$ allows shifts of *non-contiguous* spikes with no penalty.

<u>Some generalizations.</u> The families of metrics $D^{spike}[q]$, $D^{interval}[q]$, and $D^{motif}[q]$ can be generalized by modifying the cost assigned to finite translations $\Delta t$ from the simple $q(\Delta t) = q|\Delta t|$ to more general functions $Q(\Delta t)$, provided only that $Q(\Delta t_1 + \Delta t_2) \leq Q(\Delta t_1) + Q(\Delta t_2)$ (This is necessary to ensure satisfaction of the triangle inequality (1)). For example, $Q(\Delta t) = 1 - e^{-q|\Delta t|}$ is a natural choice to express a metric based on the idea that the efficacy of two spikes in driving a coincidence detector declines exponentially with their time separation (27), with rate constant $q$. Furthermore, additional metrics can be generated by combining the two or more of the steps allowed in $D^{spike}$, $D^{interval}$, and $D^{motif}$, each with their associated cost functions, to create metrics such as $D[q_{spike}, q_{interval}]$ and $D[q_{spike}, q_{interval}, q_{motif}]$.

These ideas can also be generalized to simultaneous recordings from multiple single neurons. We can regard a set of such recordings as a single spike train, in which each spike has an identified neuron of origin. This setting requires a new kind of elementary step which corresponds to relabelling the neuron of origin. In principle, the cost for this relabelling, $C(i, j)$ could depend on the neurons of origin $i$ and $j$ in an arbitrary fashion. In practice, this is likely to generate an explosion of parameters; in practice, it is likely to be sufficient simply to set $C(i, j) = C$. Values of $C$ that are small in comparison to 1 correspond to metrics which are sensitive primarily to the population firing pattern (independent of neuron of origin), while values of $C$ that are large in comparison to 1 correspond to metrics that are sensitive to the individual firing pattern of each neuron.

*Topological relationships among the metrics*

Let us now consider the extent to which the three families of metrics, $D^{spike}[q]$, $D^{interval}[q]$, and $D^{motif}[q]$, represent intrinsically different notions of distance. (In this discussion, we have chosen to implement $D^{interval}[q]$ as the variant $D^{interval:fix}[q]$, because it simplifies the analysis). That is, we ask whether closeness in the sense of one metric necessarily implies closeness in the sense of another metric. This is essentially the topological notion of "refinement:" a metric $D_a$ is said to refine a metric $D_b$ if, for every $\varepsilon > 0$, there exists a $\delta > 0$ such that

$$\text{if} \quad D_a(S, S') < \delta \quad \text{then} \quad D_b(S, S') < \varepsilon \quad . \tag{3}$$

That is, closeness in the sense of $D_a$ implies closeness in the sense of $D_b$, In other words, if metric $D_a$ refines metric $D_b$, then $D_a$ must be sensitive to all the details of temporal pattern that influence $D_b$, *provided that the spike trains that are not very different*. Moreover, if $D_a$ refines $D_b$ and also $D_b$ refines $D_a$, then the metrics are topologically equivalent (i.e., they define metric spaces that are topologically equivalent). Conversely, if $D_a$ refines $D_b$ but $D_b$ does not refine $D_a$, it is always possible to find two sequences of spike trains $S_1$, $S_2$, $S_3$,...

and $S'_1$, $S'_2$, $S'_3$,... for which the metrics $D_b(S_j, S'_j)$ approach zero, but the metrics $D_a(S_j, S'_j)$ do not. The notions of refinement and equivalence are intrinsically topological in that they are independent not only of the overall magnitude of $D_a$ or $D_b$ but also of transformations $D_i \rightarrow f_i(D_i)$ which preserve the triangle inequality (1).

For cost-based metrics, restriction of the allowed elementary steps necessarily results in a refinement of the metric. This is because placement of restrictions on (or elimination of) allowed elementary steps can never result in a smaller distance, so it suffices to take $\delta = \varepsilon$ in eq. (3). For example, $D[q_{spike}, q_{interval}]$ refines $D[q_{spike}, q_{interval}, q_{motif}]$, and $D[q_{interval}]$ refines $D[q_{spike}, q_{interval}]$.

By a similar logic, an increase in the costs of elementary steps also must result in a refinement of the metric. Thus, for $q_b < q_a$, $D^{spike}[q_a]$ is a refinement of $D^{spike}[q_b]$, $D^{interval}[q_a]$ is a refinement of $D^{interval}[q_b]$, $D^{motif}[q_a]$ is a refinement of $D^{motif}[q_b]$, and all of these metrics are refinements of $D^{count}$. This corresponds to the intuitive notion that larger values of the cost lead to greater sensitivity to the details of the temporal pattern of the spike train.

What is somewhat unexpected is that $D^{spike}[q_a]$ and $D^{spike}[q_b]$ are topologically equivalent (and similarly for $D^{interval}$ and $D^{motif}$), for any $q_a$ and $q_b$ that are nonzero. To prove this, we need to show that eq. (3) can be satisfied for sufficiently small $\varepsilon$ and $q_b > q_a$. The argument that we give for $D^{spike}[q_a]$ extends readily to $D^{interval}$ and $D^{motif}$. It suffices to consider $\delta < \min(\varepsilon q_a/q_b, 1)$. For $\delta < 1$, two spike trains $S$ and $S'$ whose distance $D^{spike}[q_a](S, S')$ is less than $\delta$ must be related by a minimal path which consists only of spike moves, since a total cost of $< 1$ excludes the possibility that any elementary step involves the insertion or deletion of spikes.  The total distance of the spike moves must be less than $\delta/q_a$, which is less than $\varepsilon/q_b$ (because of the choice of $\delta$).  Thus, the same path, viewed in $D^{spike}[q_b]$, has a cost which is no greater than $\varepsilon$.  Thus, from a topological viewpoint, the family of metrics $D^{spike}[q]$ are equivalent.  A similar conclusion holds within the family of metrics $D^{interval}[q]$ and within the family $D^{motif}[q]$ (provided $q > 0$).

However, these families of metrics are not equivalent to each other.  Rather, the metrics $D^{spike}[q]$ are all refinements of the metrics $D^{interval}[q]$, and the metrics $D^{interval}[q]$ are all refinements of the metrics $D^{motif}[q]$, but the converses are not true.  To see that the spike time metrics are refinements of the spike interval metrics, it suffices to show that *some* $D^{spike}[q_a]$ is a refinement of *some* $D^{interval}[q_b]$, because of the equivalence within each family.  It is convenient to consider $D^{spike}[q]$ and $D^{interval}[q]$.  These metrics are related because any translation of a spike by an amount $\Delta t$ can always be viewed as a change in the length of the preceding interval by $\Delta t$ and a change in the length of the following interval by $-\Delta t$. Translation of a spike by $\Delta t$ has a cost $q\Delta t$ in $D^{spike}[q]$, but the cost of the pair of changes in interval lengths in $D^{interval}[q]$ is $2q\Delta t$. This means that a path of elementary steps in $D^{spike}[q]$ can be used to generate a path of elementary steps in $D^{interval}[q]$, with at most double the cost.  Thus, it suffices to take $\delta = \varepsilon/2$ in eq. (3).  Furthermore, to see that $D^{interval}[q]$ is a refinement of $D^{motif}[q]$, one merely needs to observe that changing the length of an interval is equivalent to moving a motif consisting of all spikes which follow this interval.  That is, $D^{motif}[q] \leq D^{interval}[q]$, and one may take $\delta = \varepsilon$ in eq. (3).

To show that the spike interval metrics are *not* refinements of the spike time metrics, we display sequences of spike trains $S_1$, $S_2$, $S_3$,... and $S'_1$, $S'_2$, $S'_3$,... for which the distances $D^{interval}[q](S_j, S'_j)$ approach zero, but the distances $D^{spike}[q](S_j, S'_j)$ do not.  Let the spike trains $S_j$ consist of impulses at times 0, 1, 2, ..., $j$ - 1, $j$ and the spike trains $S'_j$ consist of impulses at times 0, 1 + 1/$j$, 2 + 1/$j$, ..., ($j$ - 1) + 1/$j$, $j$ + 1/$j$. Except for the first spike, trains differ by a displacement of 1/$j$. Thus, $D^{interval}[q](S_j, S'_j) = q/j$, with the minimal cost achieved by changing the length of the first interval. However, $D^{spike}[q](S_j, S'_j) = q$, since *each* of the last $j$ spikes must be moved by an amount 1/$j$. Thus, as $j$ increases, the distances $D^{interval}[q](S_j, S'_j)$ approach 0, but the distances $D^{spike}[q](S_j, S'_j)$ do not. Similarly, to show that spike motif metrics are not refinements of spike interval distances, we display sequences of spike trains $S_1$, $S_2$, $S_3$,... and $S'_1$, $S'_2$, $S'_3$,... for which the distances $D^{motif}[q](S_j, S'_j)$ approach zero, but the distances $D^{interval}[q](S_j, S'_j)$ do not. Here, we let the spike trains $S_j$ consist of impulses at times 0, 1, 2, ..., 2$j$ - 1, 2$j$ and the spike trains $S'_j$ consist of impulses at times 0, 1 + 1/$j$, 2 , 3 + 1/$j$, 4, ..., (2$j$ - 1) + 1/$j$, 2$j$. $D^{motif}[q](S_j, S'_j) = q/j$, which is achieved by a single step consisting of shifting all of the spikes at the odd-numbered times by an amount 1/$j$. However, $D^{interval}[q](S_j, S'_j) = 2q$, since a total of 2$j$ intervals must each be altered by an amount 1/$j$.

Despite the successive-refinement relationship of $D^{count}$, $D^{motif}$, $D^{interval}$, and $D^{spike}$, one cannot conclude that all cost-based metrics are related in a nested fashion. For example, among the variants of $D^{interval}$, one may show that $D^{interval:min}$ is topologically equivalent to $D^{interval:ign}$, and both are refined by $D^{interval:fix}$ , but only $D^{interval:fix}$ refines $D^{motif}$. This is because among the three variants of $D^{interval}$, only $D^{interval:fix}$ is sensitive to the time of the first and last spikes, and this sensitivity is needed to control $D^{motif}$. As an extreme example, consider the metrics $D^{interval:p}$, variants of $D^{interval:fix}$ in which the elementary steps include insertion and deletion of a spike, as well as shifting a contiguous group of spikes, but only if the number of shifted spikes is a power of the prime $p$. These metrics are highly unphysiological, but serve to demonstrate that it is possible to construct an infinite number of metrics, each of which is a refinement of $D^{interval:fix}$ (and hence refined by $D^{spike}$), but none of which is a refinement of any other. Similarly, the metrics $D^{motif:p}$, which allows shifting of non-contiguous subsets of spikes provided that the number of such shifted spikes shifted is a power of the prime $p$, represent an infinite number of metrics, each of which are refinements of $D^{motif}$ (and refined by $D^{spike}$), but none of which are refinements of each other. These relationships are diagrammed in Figure 1.

In sum, the notions of topological equivalence and refinement are helpful to appreciate the relationships among the metrics, considered as abstract entities. Within a class, the metrics all determine the same topological space, but different classes of metrics determine distinct topological spaces. However, as we will see below, the topological relationships do not predict which metrics lead to stronger stimulus-dependent clustering. A refinement of a given metric need not lead to stronger stimulus-dependent clustering, because the refinement may be sensitive to aspects of temporal structure that are not used by the nervous system. Additionally, clustering depends not just on the topology of the metric, but also on the relative sizes of distances between specific spike trains. Thus, we will see that although the metrics $D^{spike}[q]$ are all topologically equivalent, stimulus-dependent clustering will depend strongly on $q$.

*Efficient algorithms for the calculation of distances*

<u>Distances based on spike intervals</u>. There are simple and efficient algorithms that construct the minimal path(s) required by the definition of eq. (2) and thereby calculate the distances specified by $D^{spike}[q]$ and $D^{interval}[q]$. These algorithms are related to the elegant algorithms introduced by Sellers (37) for calculating the distance between two genetic sequences (i.e., a sequence of nucleic acid codons). For $D^{interval}[q]$, the Sellers algorithm applies directly: the spike train, considered as a sequence of interspike intervals, corresponds to a sequence of nucleotides in a DNA segment.

To compute the distance $G(E, F)$ between two spike trains whose interspike intervals are $(e_1, e_2, ..., e_m)$ and $(f_1, f_2, ..., f_n)$, we proceed inductively as follows. Define $G_{0,0} = 0$, and $G_{i,j} = 0$ for either $i$ or $j$ less than 0. Then, for $i$ or $j$ greater than 0, calculate $G_{i,j}$ as the following minimum:

$$G_{i,j} = \min \left\{ G_{i-1,j} + 1,\quad G_{i,j-1} + 1,\quad G_{i-1,j-1} + M(e_i, f_j) \right\} , \tag{4}$$

where $M(e_i, f_j)$ is the cost of changing the interval $e_i$ to the interval $f_j$, namely $q|e_i - f_j|$. Sellers (37) has shown that with this recursion rule, the distance between two subsequences $(e_1, e_2, ..., e_i)$ and $(f_1, f_2, ..., f_j)$ is given by $G_{i,j}$ . In particular, the desired distance $G(E, F)$ is given by $G_{m,n}$. Furthermore, the minimal path or paths from $E$ to $F$ are readily constructed from the options chosen at each stage of the recursion (4). The first choice corresponds to insertion of a nucleotide in sequence $E$, the second choice corresponds to insertion of a nucleotide in sequence $F$, and the third choice corresponds to changing a nucleotide. In our application, the elements (interspike intervals) form a continuum; the Sellers algorithm is concerned with sequences composed of a finite number kinds of objects (e.g., nucleotides). However, this is not crucial to the algorithm, which requires only that the rule that assigns costs to changes from one sequence element into another -- the function $M(e_i, f_j)$ -- satisfies the triangle inequality.

Distances based on spike times. The inductive idea behind the Sellers algorithm can also be used to calculate $D^{spike}[q]$, provided that the quantities $(e_1, e_2, ..., e_i)$ and $(f_1, f_2, ..., f_j)$ are considered to be spike *times* (rather than spike intervals), and the term $M(e_i, f_j)$ is $q|e_i - f_j|$, the cost of shifting a spike from time $e_i$ to time $f_j$. Despite the similarity of the algorithms for $D^{spike}[q]$ and $D^{interval}[q]$, it is somewhat awkward to prove the validity of the algorithm for $D^{spike}[q]$ from the original Sellers argument (37). It seems natural to discretize time, and then consider each spike train to be a sequence of 0's and 1's, with 0's at times without spikes, and 1's at times with spikes. But, with this formalism, a shift in time of a spike corresponds to a transposition of sequence elements, an action which is not within the realm of possibilities considered in (37).

Nevertheless, an analogous recursive algorithm is valid (47), and we sketch the argument here. Assume that we have identified a path of minimum cost $S_a = S_0, S_1, ..., S_{r-1}, S_r = S_b$ between two spike trains $S_a$ and $S_b$. The sequence of elementary steps can be diagrammed by tracing the "life history" of each spike, as shown in Figure 2. The assumption that this path is minimal places severe constraints on this diagram. The life history of each spike may consist of motion in at most one direction. Moreover, one need not consider diagrams in which a spike moves from its position in $S_a$ to an intermediate position, and then moves again to a final position in $S_b$. These constraints force one of three alternatives: either (i) the last spike of spike train $S_a$ is a spike to be deleted; or (ii) the last spike of spike train $S_b$ is a spike which is inserted; or (iii) the last spikes of both trains are connected by a shift. The validity of the recursion (4) follows directly. The similarity of the algorithms for $D^{spike}$ and $D^{interval}$ suggests that they share a common fundamental basis in the theory of dynamic programming algorithms (38), which encompasses the validation of the algorithm for $D^{interval}$ (37) and the validation of the algorithm for $D^{spike}$ (Figure 2).

Extensions. The algorithms for $D^{spike}$ and $D^{interval}$ may be readily extended to metrics in which the cost of shifting a spike (or stretching an interval) by an amount $\Delta t$ is an concave-downward function of $q\Delta t$, such as $e^{-q|\Delta t|}$. They also extend to the calculation of distances between spike trains recorded from multiple distinguished neurons, provided only that at each stage of the recursion, one adds options for relabelling the neuron of origin of the spike under consideration.

However, it is not so straightforward to extend this framework to calculate distances such as $D[q_{spike}, q_{interval}]$, which includes both elementary steps that move individual spikes, and elementary steps which change the length of interspike intervals. Two of the kinds of problems that can arise are illustrated in Figure 3. In both cases, the path of minimal total cost between spike train $S_a$ and spike train $S_b$ is achieved by a change in the length of the initial interval (the step from $S_a$ to $S$) followed by a change in the position of the spike marked * (the step from $S$ to $S_b$). In Figure 3A, this results in a "life history" for the spike marked * which includes a move to an intermediate position via a change in interval length, followed by a move to its final position via a shift in absolute time. In contrast to the situations considered by eq. (4) or illustrated in Figure 2, neither the intervals and nor the spike positions in the intermediate spike train $S$ correspond to those in either of the spike trains $S_a$ or $S_b$. A more extreme version of this difficulty is illustrated in Figure 3B. Here, provided that the number of spikes in the clusters M, $m_1$, and $m_2$ are sufficiently large, the path of minimal total cost between spike train $S_a$ and spike train $S_b$ includes a shift in the position of the spike marked * away from, and then back to, its initial position. This violates the constraint that the "life history" of each spike's movements is unidirectional. Both situations (Figure 3A and B) would not have been considered by the recursive algorithm above, which only considers life histories which are nonstop and unidirectional. Of course, this does not mean that $D[q_{spike}, q_{interval}]$ is not a metric; it merely means that the recursive algorithm might fail to find a minimal-cost path.

*Calculation of stimulus-dependent clustering*

The procedures described above for calculation of distance can be applied to any pair of neural responses. However, it is unclear to what extent these distances have any relevance to neural coding. This motivates the next step in our analysis, in which we formulate a procedure to determine to what extent the distances between individual responses (in the senses determined by the metrics $D^{spike}[q]$, $D^{interval}[q]$ ,...) depend in a systematic manner on the stimuli. This approach is intended to be applied to experimental datasets that contain multiple responses to each of several stimuli, without any further assumptions (47). If a particular metric is sensitive to temporal structure that neurons use for sensory signalling, then responses to repeated presentations of the same (or similar) stimuli should be close, while responses to presentations of distinct stimuli should be further apart. That is, in the geometry determined by the candidate metric, there should be systematic stimulus-dependent clustering.

Since we do not assume that the individual responses correspond to "points" in a vector space, we cannot use principal components analysis or other vector-space-based clustering approaches (8, 25, 26). Furthermore, even if we were able to embed the responses into a vector space, we have no guarantee that the responses to each stimulus class would lie in a blob-like "cloud"; conceivably, they could have more complex geometry, such as concentric circles. For these reasons, we seek a clustering method that makes use of nothing more than the pairwise distances themselves -- so that the identification of stimulus-dependent clustering makes a statement about the metric used to define the distances, rather than about the clustering method itself (19, 28).

More formally, we begin with a total of $N_{tot}$ spike trains, each of which is elicited in response to a member of one of the stimulus classes $s_1$, $s_2$, ..., $s_C$. We would like to use the distances between these $N_{tot}$ responses to classify them into $C$ response classes $r_1$, $r_2$, ..., $r_C$.. This classification will be summarized by a matrix $N(s_\alpha, r_\beta)$, whose entries denote the number of times that a stimulus $s_\alpha$ elicits a response in class $r_\beta$.

We proceed as follows. Initially, set $N(s_\alpha, r_\beta)$ to zero. Considering each spike train $S$ in turn, temporarily exclude $S$ from the set of $N_{tot}$ observations. For each stimulus class $s_\gamma$ we calculate $d(S, s_\gamma)$, an average distance from $S$ to the spike trains elicited by stimuli of class $s_\gamma$, as follows:

$$d(S, s_\gamma) = \left[ \left\langle \left( D[q](S, S') \right)^z \right\rangle_{S' \text{ elicited by } s_\gamma} \right]^{1/z} . \tag{5}$$

This average distance is also calculated for the stimulus class $s_\alpha$ which contains $S$, but since $S$ is temporarily excluded from the set of observations, the term $D[q](S, S)$ is excluded from (5). We then classify the spike train $S$ into the response class $r_\beta$ for which $d(S, s_\beta)$ is the minimum of all of the averaged distances $d(S, s_\gamma)$, and increment $N(s_\alpha, r_\beta)$ by 1. In the case that $k$ of the distances $d(S, s_\beta)$, $d(S, s_{\beta'})$, ... share the minimum, then each of the $N(s_\alpha, r_\beta)$, $N(s_\alpha, r_{\beta'})$, ... are incremented by $1/k$.

Note that in order to determine the average distance between a spike train $S$ and the set of responses elicited by $s_\gamma$ we have averaged the individual distances after transforming by a power law (the exponent $z$ in eq. (5)). A large negative value for the exponent $z$ would bias the average to the shortest distance between $S$ and any response elicited by $s_\gamma$, and thus would classify the spike train $S$ into the class in which there is the closest match. Conversely, a large positive value of $z$ would classify the spike train S into the class in which the distance to the furthest outlier is minimized. Not surprisingly, positive values of $z$ often lead to significantly lower estimates of the transmitted information, because of the emphasis on distances from the outliers.

$N(s_\alpha, r_\beta)$ is the number of times that a stimulus from class $\alpha$ is classified as belonging to class $\beta$. If

this classification were perfect, then $N(s_\alpha, r_\beta)$ would be diagonal. If this classification were random, then the on- and off-diagonal elements of $N(s_\alpha, r_\beta)$ would be comparable. We use an information-theoretic measure, the transmitted information (4), to quantify the extent to which this classification is non-random. For stimuli that are drawn from discrete classes $s_1$, $s_2$,..., and spike train responses that have been grouped into discrete classes $r_1$, $r_2$, ...., the transmitted information $H$ is given by

$$H = \frac{1}{N_{tot}} \sum_{\alpha,\beta} N(s_\alpha, r_\beta)\left[ \log_2 N(s_\alpha, r_\beta) - \log_2 \sum_a N(s_a, r_\beta) - \log_2 \sum_b N(s_\alpha, r_b) + \log_2 N_{tot} \right] . \qquad (6)$$

For $C$ equally-probable stimulus classes, random classification corresponds to $N(s_\alpha, r_\beta) = N_{tot}/C^2$ and yields $H = 0$. Perfect classification (which corresponds to $N(s_\alpha, r_\beta) = N_{tot}/C$ when $\alpha = \beta$, and 0 otherwise) yields a maximal value of the transmitted information $H$, namely $\log_2 C$. In principle, this maximum might also be achieved by a matrix $N(s_\alpha, r_\beta)$ for which $N(s_\alpha, r_\beta) = N_{tot}/C$ when $\alpha = P(\beta)$ (for $P$ a nontrivial permutation). This corresponds to a situation in which clustering is perfect, but the classification algorithm mislabels the response classes. This is very unlikely to arise, given the structure of our classification scheme, since every response would have to be closer to responses to stimuli from other classes, than to responses from its own class.

It is worth emphasizing that in using an information-theoretic measure of clustering, we do not mean to imply that the quantity $H$ represents the information available to the nervous system, or that the analytical stages we describe here have any correspondence to neural processing of information. Rather, this choice is simply intended to be a measure of clustering, which does not make assumptions about parametric relationships, if any, among the stimulus classes.

Biases due to small sample size. If the number of presentations of each stimulus class is small, then the value of $H$ estimated by eq. (6) will be spuriously high, simply because of chance proximities between the few examples of observed responses. Even if there are a large number of stimulus presentations, one anticipates an upward bias of the estimate of $H$. For example, in the case of $C$ equally probable stimulus classes, any deviation of $N(s_\alpha, r_\beta)$ from the expected value of $N_{tot}/C^2$ will result in a positive value of the estimated transmitted information $H$. This problem represents a general difficulty in the estimation of transmitted information from limited samples (7).

Recently, Treves and Panzeri (45) have derived an analytic approximation to this bias: asymptotically, the bias (for a fixed number of stimulus and response bins) is inversely proportional to the number of samples and is independent of stimulus and response probabilities. The derivation of this asymptotic behavior assumes that there is a binning process which treats all responses in an independent manner. The present approach explicitly avoids such a binning process, so there is no guarantee that a similar correction is applicable. For this reason, we choose a computational approach to estimate this bias (9, 29): we use eq. (6) to recalculate the transmitted information $H$ following random reassignments of the observed responses $S$ to the stimulus classes. Values of the un-resampled $H$ which lie in the upper tail of this distribution are thus unlikely to represent a chance grouping of responses. Furthermore, the average value of many such calculations, which we denote $H_0$, is an estimate of the upward bias in the estimate of $H$.

*Examples: simple simulations*

We now present some examples in which we apply the above procedures to some simple numerical simulations. These numerical simulations are not intended to be realistic, but rather to illustrate some of the behaviors described above.

Rate discrimination: regular and irregular trains. The first example (Figure 4) concerns stimuli which elicit different mean rates of firing. We considered five stimulus classes, each of which elicited firing at different average rates ($R$ = 2, 4, 6, 8, and 10 impulses/sec). Twenty examples of one-second responses to each stimulus were simulated, and the transmitted information was calculated from these $N_{tot}$= 100 spike trains according to eq. (6). These calculations were repeated for forty independent synthetic datasets, to determine the reliability (± 2 s.e.m.) of these estimates of transmitted information $H$. For each of these forty datasets, two resamplings by relabelling, as described above, was performed to estimate the contribution $H_0$ due to chance clustering. These calculations were performed for values of $q$ spaced by factors of 2, and for four values of the clustering exponent $z$: -8, -2, 2, and 8. We focus on the behavior of $D^{spike}$; the behavior of $D^{interval}$ was very similar (data not shown).

In Figure 4A-D, the spikes were distributed in a Poisson fashion (c.v. = 1.0) for each of the five firing rates $R$. In this case, the precise time of occurrence of each spike carries no information. Indeed, more clustering (higher $H$) is seen for $D^{count}$ ($D^{spike}[0]$) than for $D^{spike}[q]$, for $q > 0$. For low values of $q$, the decrease in $H$ for $q > 0$ is relatively minor (from about 0.7 to 0.5 for negative $z$, and essentially no decrease for positive $z$). In this range ($q < 16$), $D^{spike}[q]$ is influenced by a mixture of spike counts and spike times. For sufficiently high values of $q$, $H$ falls to chance levels. In this range ($q > 32$), the cost of moving a spike is sufficiently high so that nearly all minimal paths correspond to deletion of all spikes followed by reinsertion at other times. Consequently, the defining feature of each stimulus class (i.e., a similar number of spikes) is ignored, and responses within the same stimulus class, especially at high firing rates, are seen (by $D^{spike}[q]$) as very different.

Figure 4E-H shows a similar calculation, but with a spike generating process that is more regular (c.v. = 0.125). To simulate responses to a class characterized by a steady mean rate $R$, we selected every $k$th spike from an underlying Poisson process with rate $kR$. (The first spike is chosen at random from the initial $k$ spikes of this underlying process). In this "iterated Poisson" process, the interspike intervals were distributed according to a gamma distribution of order $k$:

$$p(t) = \frac{1}{\Gamma(k)} R^k t^{k-1} e^{-Rt} \quad ,$$

where $p(t)$ is the frequency of interspike intervals of length $t$. In the simulations shown here, we chose $k = 64$, and thus the interspike intervals have a coefficient of variation of 0.125 ($=1/\sqrt{k}$). Because the spike trains are regular, there is less variability in the number of spikes in each sample. Consequently, the spike count metric $D^{count}$ leads to a greater degree of clustering than in the Poisson case (Figure 4A-D) -- calculated values of $H$ are approximately 2.0, close to the maximum achievable value of $\log_2(5) \approx 2.32$. Nevertheless, for small values of $q$ and negative values of $z$ (Figure 4E-F), there is a further increase in $H$. This increase reflects the fact that within a class, responses will be similar not only in the number of spikes: additionally, responses whose first spikes occur at similar times will have subsequent spikes that occur at similar times. That is, the optimal match between a response to a given stimulus and other responses in the same class will be stronger if spike times, and not just spike counts, are considered. However, there is an optimal choice for $q$, beyond which further increases lead to decreasing values of $H$. This is because a cost $q$ corresponds to the notion that the timing of a spike matters up to an amount $1/q$. Choosing $q$ too high for the typical precision of the spike times causes the distance measure to be influenced by details that are irrelevant to the response classes. Consequently, for high $q$, the apparent clustering decreases.

Note that for positive $z$ (Figure 4G-H), there is no increase in $H$ for any value of $q$. This is because the clustering based on the average distance as defined by eq. (5) now assigns each response to the class with the "least bad" match, rather than the "most good" match. Sensitivity to spike times improves the quality of the best match, but worsens the quality of the worst match.

We also point out that, despite the fairly large dependences $H$ on $q$ and $z$ seen in Figure 4, there is relatively little dependence of $H_0$, the estimate of chance information. That is, the increase in temporal resolution associated with high $q$ is not associated with large upward biases in estimates of information, as would have been the case if increased temporal resolution were to require increasingly sparse sampling of a space of progressively higher dimension.

This simulation shows that for response classes which differ in overall firing rate (and thus, are readily discriminable by a simple rate code), metrics which are sensitive to temporal pattern allow enhanced discrimination. This modest enhancement is restricted to regular spike trains (Figure 4E-H) and values of $q$ which are not so large as to regard even minor shifts of spike times as "different." For irregular trains, and for high values of $q$, attention to temporal structure reduces discrimination, as would be expected since the temporal structure is irrelevant to the response classes.

Temporal phase discrimination. The next three examples present situations in which $D^{spike}$ and $D^{interval}$ have contrasting behavior. Figure 5 shows a simulation of temporal phase discrimination. There were four stimulus classes, each with 20 simulated responses of 1 second duration. In Figure 5A-D, we used a time-dependent Poisson process to generate spike trains. The instantaneous firing density $R(T)$ was given by $R_0[1 + m\cos(2\pi fT + \phi)]$, where the mean firing rate $R_0$ was 20 impulses/sec, the modulation frequency $f$ was 4 Hz, and the modulation depth $m$ was 0.5. The stimulus classes differed in their modulation phase $\phi$, which was chosen from the set $\{0, \pi/2, \pi, 3\pi/2\}$. Thus, responses to different stimulus classes were anticipated to differ in the arrangement of spike times, and not in the average total number of spikes (or in the distribution of the number of spikes). The information-theoretic index of clustering $H$, and the clustering $H_0$ due to chance alone, were calculated from 40 independent simulations, as in Figure 4. As is seen in Figure 5A-D, clustering was at chance level for $D^{count}$. In contrast, $D^{spike}[q]$ revealed highly significant clustering. The peak of $H$ at $q = 32$ corresponds to 1/8 of a cycle (half the phase increment which separates the stimulus classes). For substantially smaller values of $q$, clustering is diminished because the distinction between the classes is diminished; for substantially larger values of $q$, clustering is diminished because the distance is influenced by jitters in the spike times which are irrelevant to the response classes. This behavior was seen for all values of the clustering exponent $z$.

Figure 5A-D also shows the degree of clustering revealed by $D^{interval:fix}[q]$ and $D^{interval:min}[q]$. These metrics, which are sensitive to the sequence of interspike intervals but not to the absolute time of occurrence of spikes, reveal almost no clustering beyond chance. The two metrics differ only in how they treat the first

and last interspike intervals (i.e., the time between the start of the response and the first spike, and the time between the last spike and the end of the response period): $D^{interval:fix}[q]$regards these intervals as fixed, while $D^{interval:min}[q]$ adjusts them to minimize the distance. As such, $D^{interval:fix}[q]$ would be expected to retain a slight sensitivity to temporal phase, because it retains the first and last intervals. However, this only results in a minimal (<0.05) increase in $H$.

In Figure 5E-H, the same instantaneous firing densities $R(T) = R_0[1 + m \cos(2\pi fT + \phi)]$ were used to drive an iterated Poisson process of order $k = 64$. The resulting spike trains are much more regular (eightfold decrease in the coefficient of variation of the interspike interval), and clustering is much stronger. Furthermore, the range of values of $q$ for which $D^{spike}$ shows significant clustering is substantially extended. The range of effective values of $q$ extends further downward because shifts in time of multiple adjacent spikes are correlated. The range of effective values of $q$ extends further upward (approximately by a factor of 8) because of the proportionate decrease in the coefficient of variation and the corresponding shortening of the timescale at which precise timing becomes meaningless. Somewhat surprisingly, $D^{interval:fix}[q]$ and $D^{interval:min}[q]$ also reveal significant clustering for these spike trains. This is an indirect consequence of the regularity of the interspike intervals. One candidate path of elementary steps between responses from different stimulus classes will consist of deleting some interspike intervals from the beginning of one response, deleting some interspike intervals from the end of the other response, and stretching the intervening intervals to achieve a match. This kind of path will reveal phase-dependent clustering. However, whether this path is optimal (and hence, whether its total cost determines the distance) depends on the tightness of the match between interspike intervals, and thus, on regularity of the spike trains. $D^{interval:fix}[q]$ has an advantage over $D^{interval:min}[q]$ because of the end-effect discussed above provides independent absolute phase information, but this advantage is slight. All of these findings are largely independent of the choice of the clustering exponent z.

In sum, this simulation shows that $D^{spike}$, but not $D^{count}$ or $D^{interval}$, can discriminate among irregular spike trains which vary in temporal phase (Figure 5A-D). For regular spike trains with instantaneous firing frequencies, both $D^{interval}$ and $D^{spike}$ can perform the discrimination (Figure 5E-H).

Temporal frequency discrimination. In this simulation (Figure 6), there were four stimulus classes, each with 20 simulated responses of 1 second duration, and an instantaneous firing density $R(t)$ was given by $R_0[1 + m\cos(2\pi ft + \phi)]$, with a mean firing rate $R_0$ of 20 impulses/sec. For three of the classes, the modulation depth $m$ was 0.5, and the phase $\phi$ was chosen at random in each trial. These classes were characterized by the modulation frequency $f$, which was chosen from the set {3, 4, 5}. The fourth class was unmodulated ($m = 0$, $f$ and $\phi$ irrelevant). Thus, responses to different stimulus classes were anticipated to differ primarily in the arrangement of interspike intervals (because they differed in modulation frequency), but not in absolute spike times (since the initial phase $\phi$ was chosen at random). The information-theoretic index of clustering $H$, and the clustering $H_0$ due to chance alone, were calculated as in Figure 4 and Figure 5. For spike times generated by a time-dependent Poisson process (Figure 6A-D), there is essentially no stimulus-dependent clustering for any of the metrics considered. This is because the irregularities of the interspike interval distributions due to the random times of occurrence of the spikes dominate the systematic temporal modulation. However, for spike times generated by the more regular iterated Poisson process (Figure 6E-H), substantial stimulus-dependent clustering is seen. We first consider $D^{interval:fix}[q]$ and $D^{interval:min}[q]$. There are two regimes: for $q$ greater than 0 but no greater than 8, clustering is substantially greater than chance, but not sufficiently great to reliably distinguish the four classes from one another. In this range, the degree of clustering depends strongly on the clustering exponent $z$ (increasing as $z$ decreases), indicating that it reflects the presence of a few good matches rather than the absence of bad matches. For larger values of $q$, the degree of clustering approaches the maximal possible value of $\log_2(4) = 2$. The height and position of this peak (though not its breadth) are relatively independent of $z$. $D^{interval:min}[q]$ has a slight advantage over $D^{interval:fix}[q]$, corresponding to the fact that it ignores the initial and final interspike intervals, which are irrelevant to the distinction between the classes. Note that although these stimulus classes were constructed to be distinguished on the basis of spike intervals, rather than spike times, $D^{spike}[q]$ nevertheless reveals significant clustering. This is analogous to the behavior of $D^{interval:fix}[q]$ and $D^{interval:min}[q]$ in phase discrimination (Figure 5E-H). However, clustering in the sense of $D^{spike}[q]$ is sensitive to $z$ because good matches at one relative phase will necessarily imply bad matches at other relative phases. Thus, average distances (as defined by eq. (5)) that are weighted heavily by the best match (i.e., $z < 0$) will tend to reveal the greatest degree of clustering.

<u>Identification of deterministic chaos</u>. In the next simulation (Figure 7), the two response classes are distinguished by the presence or absence of deterministic chaos (loosely inspired by observations of low-dimensional chaos in the olfactory system (41)). In both stimulus classes, responses consisted of impulse trains in which the interspike intervals were uniformly distributed between 0 and $2/R_0$, with the mean rate $R_0$, set at 10 impulses/sec. For spike trains in the first class, these interspike intervals were placed in random order. For spike trains in the second class, the first interspike interval is chosen at random, but subsequent interspike $I_{n+1}$ intervals are determined from the preceding interval $I_n$ by the Baker transformation,

$$I_{n+1} = \begin{cases} 2\, I_n\,, & I_n \le \dfrac{1}{R_0} \\ 2\left(\dfrac{2}{R_0} - I_n\right), & I_n > \dfrac{1}{R_0} \end{cases} . \qquad (8)$$

For each calculation of $H$, 100 1-second examples of each response were generated. $H$ and $H_0$, and their s.e.m.'s, were estimated as in the previous simulations from 20 independent simulations. These spike trains had identical distributions of interspike intervals, identical mean rates ($R_0$ = 10 impulses/sec), flat post-stimulus time histograms, and no pairwise correlations between intervals at second order. Thus, in contrast to the previous simulations (Figure 4, Figure 5, and Figure 6), traditional approaches such as Fourier analysis, as well as some more recently-proposed ideas (13, 26) would not have been able to distinguish these response classes.

As seen in Figure 7, substantial clustering beyond that expected from chance alone is seen for all three metrics considered. The advantage of $D^{interval:min}[q]$ and $D^{interval:fix}[q]$ over $D^{spike}[q]$ makes sense, because of the interval-based nature of the temporal structure. Interestingly (Figure 7A-B), in contrast to the previous simulations, some estimates of $H$ continue to rise for high values of $q$. This is because successive iterations of the transformation (eq. (8)) are sensitive to indefinitely small changes in interspike interval lengths, as is characteristic of chaotic processes in general. Thus, sensitivity to tiny changes in interval length improves the quality of mutual matches among samples of the chaotic process. This behavior is not seen for $D^{spike}[q]$ (which is only indirectly sensitive to interval length), nor for positive values of $z$ (where clustering is weighted by the absence of mismatches, rather than by the closeness of the best match).

*The geometries induced by a family of metrics*

We now return to a simple rate-discrimination simulation to illustrate the way that changes in the metric induce changes in the nature of clustering. We consider two stimulus classes, which elicit responses of mean rates 6 and 7 impulses/sec respectively and have interspike intervals determined by an iterated Poisson process of order $k = 64$. For this simulation, 40 example spike trains for each class were simulated, and the information-theoretic measure of clustering, $H$, was calculated for $D^{spike}[q]$ for a range of values of $q$ and a clustering exponent $z = -2$. The dependence of $H$ on $q$ (Figure 8) was similar to that seen in Figure 4E-H: a modest increase in clustering for $q > 0$, followed by an abrupt fall-off for values of $q$ sufficiently high as to force sensitivity to irrelevant detail in the spike trains. However, despite the modest change in $H$, there is a qualitative change in the nature of the clustering for low values of $q$ (e.g., $q = 1$), and values of $q$ near the peak in $H$ (e.g., $q = 16$). To see this, we use multidimensional scaling (16, 20) to embed the spike trains into a Euclidean space. This procedure assigns coordinate $n$-tuples to each spike train, so that the standard Euclidean distances between these $n$-tuples are as close as possible to the distances yielded by a given metric, in this case $D^{spike}[q]$. Successive coordinates correspond to eigenvectors of a symmetric matrix $M_{jk}$, scaled by the square roots of the corresponding eigenvalues. The entries of the matrix $M_{jk}$ are given by

$$M_{jk} = -\frac{1}{2}\left( D_{jk}^2 - \langle D_{jr}^2 \rangle_r - \langle D_{rk}^2 \rangle_r + \langle D_{rs}^2 \rangle_{rs} \right) \quad , \qquad (9)$$

where $D_{jk}$ indicates the distance between spike train $j$ and spike train $k$, $<\ >_r$ indicates an average over all spike trains $r$, $<\ >_s$ indicates an average over all spike trains $s$, and $<\ >_{rs}$ indicates an average over all pairs of spike trains $r$ and $s$. Note that this embedding does not guarantee that all points in the Euclidean space correspond to spike trains -- merely that the distances of spike trains, in the sense of $D^{spike}[q]$, are well-approximated by the Euclidean distances between their $n$-tuples. Additionally, the matrix of eq. (9) may have negative eigenvalues, corresponding to a hyperbolic geometry in the embedding space. Nevertheless, the embedding provides a visual way to understand the geometry induced by the metrics $D^{spike}[q]$.

The multidimensional scaling procedure was applied to each response class (40 samples) in isolation, as well as to the combined dataset of 80 samples. Figure 9A shows the results for $q = 1$. Consider first the multidimensional scaling of the "6 imp/sec" class in isolation (left inset). For $q = 1$, $D^{spike}[q]$ is determined primarily by the number of spikes. Most spike trains in the "6 imp/sec" class have 6 spikes, but some have 5 spikes, and some have 7 spikes. Hence, the spike trains form three clouds, arrayed approximately in a line, one corresponding to each spike count. Spike trains with the same number of spikes are close to each other but not identical. This is reflected in the dispersion of the points within each cloud and the requirement for more than one dimension to account fully for the distances determined by $D^{spike}$. A similar picture is seen for the multidimensional scaling of the "7 imp/sec" class in isolation (right inset). Combined multidimensional scaling (main scattergram) of the two classes yields four clouds, corresponding to the four possible spike counts of 5, 6, 7, and 8 impulses. Only the "6 imp/sec" class contributes to the cluster corresponding to 5 impulses (the three left-most clusters), and only the "7 imp/sec" class contributes to the cluster corresponding to 8 impulses (the three right-most clusters), and both stimulus classes contribute to the clusters with 6 and 7 impulses. This overlap corresponds to the imperfect discrimination of these two classes by clustering based on $D^{spike}[1]$. The scatter along the second dimension does not help to disambiguate these responses.

Figure 9B shows the corresponding picture for $q$ = 16. $D^{spike}[16]$ is sensitive to times on the order of 0.1 sec, and on this scale, both spike trains are quite regular (mean intervals of 0.166 and 0.143 sec, with c.v. of 0.125). The locus corresponding to the "6 imp/sec" class in isolation (left inset) now forms an approximate circle. This corresponds to the fact that for $D^{spike}[16]$, the distance between two fairly regular spike trains is determined primarily by their relative phase. Multidimensional scaling of the "7 imp/sec" class in isolation gives a similar picture (right inset). Note that there are no points at the center of either ring: such points would correspond to spike trains which are not only equidistant from all of the spike trains, but also close to all of them; such spike trains (in the context of $D^{spike}[16]$) do not exist. Multidimensional scaling of the combined classes requires four dimensions -- the first four eigenvalues of the matrix of eq. (9) are 0.348, 0.284, 0.240, and 0.208. In the two-dimensional projection illustrated (main scattergram), the points corresponding to the two stimulus classes appear to form interpenetrating clouds. Examination of the higher-dimensional representation reveals that these clouds correspond, approximately, to two mutually orthogonal circles, one corresponding to each class. That is, within each class, the spike trains' locus is approximately circular, and all points in one class are, approximately, at the same distance from all points in the other class. In contrast to the situation for $q$ = 1, multiple dimensions contribute to separation of the response classes. Furthermore, the geometrical center of the two response classes is similar -- near the origin -- and hence, a clustering scheme which assumed that the responses lay in convex clouds would have failed.

Yet a third kind of behavior is seen for $D^{spike}[256]$. This distance is sensitive to times on the order of 0.01 sec. On this timescale, both classes of spike trains are irregular. Consequently (Figure 9C), no discernable geometrical structure is apparent in the multidimensional scaling of the spike trains in isolation (left and right insets): they form what appears to be a random cloud. Multidimensional scaling of the combined classes again forms a random cloud (main scattergram), either when inspected in the projection illustrated, or in projections along higher-order eigenvectors. This corresponds to the fact that this metric is sensitive to differences between individual responses (idiosyncratic timing of spikes) which are unrelated to differences between the stimulus classes.

These three regimes are summarized by the analysis of Figure 10. We introduce a dimension index *E*, which describes the effective dimension of an embedding. The dimension index *E* is defined by

$$E = \frac{(\sum \lambda_i)^2}{\sum \lambda_i^2} , \tag{10}$$

where $\lambda_i$ are the eigenvalues of the matrix of eq. (9). Note that for an embedding in which each of $n$ dimensions contributed equally ($\lambda_i = 1/n$), this index would have the value $n$. For $q$ in the range 4 to 32, the dimension index for the embedding of each class in isolation is approximately 2 (light and dark symbols), and the dimension index for the combined classes is approximately 4 (solid line without symbols). This corresponds to a circular locus for each of the classes in isolation, and bi-orthogonal circles for the embedding of the combined classes (Figure 9B). For lower values of $q$, the effective dimension decreases towards 1, and the dimension index for the embedding of the combined classes is no larger than that of the classes in isolation. That is, the higher dimensions do not contribute to stimulus-dependent clustering. For values of $q > 32$, the dimension index rises rapidly, but, as we have seen in Figure 9C, this increase in dimension does not produce any discernable structure. It should be noted that one can always achieve an embedding of $n$ points in a space of dimension $n$-1 (39 and 79 in the present case), and thus (for $q > 128$) it is likely that our estimates of dimension in Figure 10 are lower bounds, limited by the number of spike trains we chose to generate.

However, it is not the case that a progressive increase in embedding dimension necessarily implies a decrease in clustering. Indeed, the opposite situation is seen for the discrimination of deterministic chaos and randomly-sequenced interspike intervals. Figure 11 shows the dependence of the dimension index $E$ (eq. (10)) on $q$ for embeddings of the chaotic spike trains in isolation, the random spike trains in isolation, and the two classes combined. For all three families of metrics (Figure 11A: $D^{spike}$; Figure 11B: $D^{interval:fix}$; Figure 11C: $D^{interval:min}$), the set of chaotic spike trains is associated with a lower embedding dimension than the random trains, and this dimension increases monotonically with $q$. In contrast to the simple geometrical situation of the regular spike trains (Figure 9C), the rise in dimension is not associated with a decrease in clustering -- rather, as we have seen in Figure 7, clustering continues to rise with increasing $q$. It is also worthwhile to note that the high embedding dimension is not associated with an increase in the chance clustering (dashed lines in Figure 7). This suggests that our purely metrical approach to clustering has successfully circumvented the problem of attempting to perform clustering in a sparsely-populated high-dimensional vector space, which would be anticipated to lead to progressively higher estimates of chance clustering with increasing dimensionality.

*Comparison with a benchmark set of simulated data*

The final simulation utilizes a benchmark dataset, developed by Golomb et al. (14) to compare a variety of methods to determine the information content of spike trains. These simulations (15) generate responses of idealized lateral geniculate magnocellular and parvocellular neurons to spatial patterns consisting of Walsh functions. The responses to each Walsh pattern are derived from a time-dependent Poisson process, whose envelope is determined by convolving the Walsh pattern with modelled spatiotemporal filtering properties of lateral geniculate neurons. For the simulations of Figure 12A and B, datasets are constructed from 64 responses (250 msec in duration) to each of the 32 patterns, and the above procedures are used to calculate information-theoretic measures of clustering $H$ for $D^{spike}[q]$ and $D^{interval:min}[q]$. $H_0$ is determined by repeating this calculation for 5 random reassignments of the responses to stimuli. This procedure is repeated for 16 independent datasets of 64 responses per pattern.

As seen in Figure 12A and B, the estimate of chance clustering, $H_0$, is substantially larger than in the simulations discussed previously, as one would expect from the fact that there is a relatively small ratio of samples to classes (7, 45). For the modelled magnocellular neuron (Figure 12A), clustering is maximal for $D^{spike}[q]$ with $q_{max} = 32$. The increase in $H$ for $D^{spike}[32]$ compared to $D^{count} = D^{spike}[0]$ exceeds the corresponding change in $H_0$. That is, the measure of clustering corrected for the bias due to small sample size, $H - H_0$, is largest for $D^{spike}[32]$. For $q < 128$, clustering for $D^{spike}$ exceeds clustering for $D^{interval:min}$, whether or not the correction term $H_0$ is subtracted. Thus, for the magnocellular neuron of this simulation, the timing of impulses (to within a precision of ca. $1/q_{max} = 30$ msec) conveys significant stimulus-dependent information. For the modelled parvocellular neuron (Figure 12B), the situation is quite different. There is only a modest increase in $H$ for $q > 0$, and this increase is less than the increase in $H_0$. That is, $H - H_0$ is largest for $D^{count}$; there is no evidence for improvement in clustering either for $D^{spike}$ or $D^{interval:min}$.

Figure 12C and D show how this analysis depends on the number of samples per stimulus class. As the number of samples increases, $H$ and $H_0$ decrease. For each metric considered, these decreases are approximately parallel, and consequently, the corrected estimator of clustering, $H - H_0$. changes by a relatively small amount. (Nevertheless, it is notable that this drift is upward, which suggests that subtraction of $H_0$ may be an overly conservative correction.) For the modelled magnocellular neuron (Figure 12C), $H - H_0$. is largest for $D^{spike}$ provided that there are at least 32 samples per stimulus class. For the modelled parvocellular neuron (Figure 12D), $H - H_0$. is largest for $D^{count}$ for all values of the number of samples per stimulus class.

It is interesting to compare our empirical estimates of $H_0$. with the analytic result of Treves and Panzeri (45). Consistent with this result, there is an approximate inverse proportionality between $H_0$ (dashed lines in Figure 12C and D) and the number of samples per stimulus class. However, this proportionality constant is dependent on the metric and also (dashed lines in Figure 12A and B) on the value of $q$. This appears to be at variance with the analytic result (45), which states that asymptotically, this proportionality constant depends only on the size of the stimulus and response classes, and the number of samples. However, there is no contradiction: Treves and Panzeri's derivation (45) requires that the assignments of samples to response categories be made independently. Independence is violated by the clustering scheme we used, and the consequent dependence of the proportionality constant on the metric that we observe emphasizes the necessity of the assumption of independence for the analytic result (45).

The estimates in Figure 12C and D of $H$ and $H - H_0$ may be compared directly with the calculations of Golomb and coworkers (14) for these simulated datasets, based on a binning approach and on neural network classifiers (19). For the magnocellular neuron, our estimate of $H - H_0$ for $D^{count}$ (0.19 for 128 samples/class) approximated the estimates obtained by Golomb et al. (14) (≈ 0.24 for 128 samples/class for binning or for network classifiers) for the "number of spikes" code. We found an increase in $H$ - $H_0$ to 0.34 for $D^{spike}$, while Golomb et al. (14) found an increase (to ≈ 0.5 for the binning method, ≈ 0.4 for the neural network classifier method) for multidimensional codes based on three principal components. For the parvocellular neuron, our estimate of $H$ - $H_0$ for $D^{count}$ (0.81 for 128 samples/class) approximated the estimates obtained by Golomb et al. (14) (≈ 0.85 for 128 samples/class for binning or neural network classifiers) for the "number of spikes" code. We found no increase in $H$ - $H_0$ for $D^{spike}$ or $D^{interval:min}$; Golomb et al. (14) found a slight increase (≈ 0.95 for binning or neural network classifiers) for multidimensional codes based on three principal components.

*Some neurophysiological examples*

In Figure 13, we re-analyze data recently published by Middlebrooks et al. (28) as evidence for temporal encoding of the azimuth of an auditory stimulus. The spike trains were recorded from a single neuron in cat anterior ectosylvian cortex following noise bursts of 100 msec duration which were presented in 20 deg steps of azimuth in the horizontal plane. For the analysis presented here, the 10 responses to each of these 18 locations were rebinned into sets of 30 responses corresponding to 60 deg sectors. Figure 13A shows the information-theoretic measure of clustering, $H$, for $D^{spike}$ and $D^{interval:min}$, along with values of $H_0$ as calculated from 10 random permutations of the data. There is a clear rise in $H$ for $q > 0$, both for $D^{spike}$ and $D^{interval:min}$. The increase for $D^{interval:min}$ seems somewhat greater than for $D^{spike}$ , but there is a similar difference in the correction term $H_0$. Additionally, there is a sharp peak in $D^{spike}[q]$ near $q = 256$, which is not associated with an increase in the correction term $H_0$. This indicates that spike times, up to a precision of $1/q \approx 4$ msec, cluster in a stimulus-dependent fashion. (Note that the ±2 s. e. m. lines in Figure 13A represent are confidence limits for the mean of $H_0$, and thus are only appropriate for comparison among different resampled estimates. To estimate the probability that the un-resampled value of $H$ could have come from this population, a ±2 standard deviation criterion ($\sqrt{10} \approx 3.2$ times the s. e. m) should be used; by this measure, $H$ is significant for all values of $q$.)

In Figure 13C and D, we present the results of multidimensional scaling of these responses, with the distance function provided by $D^{spike}[256]$. Figure 13C shows the loci which correspond to individual responses, as projected into the plane of the first two eigenvalues. There is certainly substantial overlap of the clouds corresponding to each azimuth range, but nevertheless, a trend towards segregation of responses from each sector is apparent. This trend is made much more vivid in Figure 13D, which shows only the geometrical center of each cloud. The dimension index $E$ (eq. (10)) of this embedding is 2.8, and the first 10 eigenvalues are shown in Figure 13B. The first four eigenvalues are positive, but projections of the loci of the spike trains on the third and fourth axes do not reveal additional geometrical structure.

In Figure 14, we analyze a dataset obtained in our study (47) of temporal coding in macaque visual cortex. The data consist of activity recorded in primary visual cortex in an awake animal performing a fixation task. Stimuli consisted of transient presentation of gratings of five spatial frequencies (1, 3, 5, 11, and 21 c/deg), aligned in each of 8 orientations (0 deg to 157.5 deg in 22.5 deg steps), at a contrast of 1.0 (Here, contrast is defined as $[(L_{max} - L_{min})/(L_{max} + L_{min})]$, where $L_{max}$ and $L_{min}$ are the maximum and minimum of the stimulus luminance.) For each of these 5 x 8 = 40 stimuli, 16 responses of 256 msec were collected and analyzed. As seen in Figure 14A, there is temporal coding both in the sense of $D^{spike}[q]$ and $D^{interval:min}[q]$, as shown by the rise in $H$ for $q > 0$. However, there is a falloff in $H$ by $q \approx 50$, indicating a much looser jitter of $1/q \approx 20$ msec for the temporal structure in this neuron's encoding in comparison with the auditory neuron of Figure 13. Multidimensional scaling of the responses according to the metric $D^{spike}[32]$ requires at least three dimensions to reveal the geometrical structure of the response loci (Figure 14C, D, and E). For each of the four lowest spatial frequencies, the responses to different orientations form a loop, and these loops are largely non-overlapping when examined in three dimensions. For example, the orbits for spatial frequencies 3 and 4 overlap in dimensions 1 and 2, but are well-separated in dimension 3. Conversely, the orbit for spatial frequency 1 nearly collapses in dimensions 1 and 3, but is clearly delineated along dimension 2. The dimension index $E$ (eq. (10)) of this embedding is 1.8, and the first 10 eigenvalues are shown in Figure 14B. Projections along combinations of the third and fourth axes provide additional separation of the orbits, but do not reveal qualitatively new structure.

In Figure 15, we analyze activity of a complex cell in the supragranular layers of primary visual cortex, recorded in an anaesthetized, paralyzed macaque. (Physiological preparation for recordings in the anaesthetized, paralyzed macaque are described in (48)). Stimuli consisted of transient presentation of gratings of three spatial frequencies (0.5, 2, and 4 c/deg) and five contrasts (0.0625, 0.125, 0.25, 0.5, 1.0), aligned at the optimal orientation. For each of these 3 x 5 = 15 stimuli, 128 responses of 256 msec were collected and analyzed. Again there is temporal coding both in the sense of $D^{spike}[q]$ and $D^{interval:min}[q]$, as shown by the rise in $H$ for $q > 0$. As in data of Figure 14, multidimensional scaling of the responses (Figure 15C, D, E) for $D^{spike}[64]$ reveals a systematic separation of the loci for each of the three spatial frequencies and the five contrasts, but three dimensions are needed to see this separation clearly. Responses to the lowest spatial frequency are well-separated from the responses to the two higher spatial frequencies in the projection on dimensions 1 and 2, while responses to the two higher spatial frequencies are best separated from each other in the projection on dimensions 1 and 3.

In contrast to the data of Figure 14, clustering is greater for $D^{interval:min}[q]$ than for $D^{spike}[q]$ at large values of $q$. However, multidimensional scaling (Figure 16B, C, D) for $D^{interval:min}[64]$, the same value of $q$ as in Figure 15, does not provide a systematic separation of the loci for each of the three spatial frequencies. Rather, the loci for all stimuli are nearly collinear, and the loci for the two higher spatial frequencies overlap in all projections. This qualitatively simpler structure for the geometry induced by $D^{interval:min}[q]$ is made quantitative by the greater dominance of the first eigenvalue (Figure 16A compared with in Figure 15B), and the smaller dimension index $E$ (1.6 for $D^{interval:min}[64]$, compared with 3.7 for $D^{spike}[64]$). That is, even though $D^{interval:min}[64]$ provides a greater degree of clustering than $D^{spike}[64]$, $D^{spike}[64]$ induces a metrical structure which separates the two stimulus attributes of contrast and spatial frequency, while $D^{interval:min}[64]$ does not.

# DISCUSSION

*Overview*

We have presented a new approach to the analysis of the temporal structure of spike trains. The notion of "distance" between spike trains plays a central role in this approach. Based solely on the notion of distance, we show how it is possible to assess (via an information-theoretic measure) the extent to which experimentally-measured neural responses cluster in a systematic fashion. This measure of clustering indicates to what extent this candidate distance is sensitive to features of spike trains that convey stimulus-specific information. By repeating this process for several families of distances, one can learn about the nature and precision of temporal coding.

The notions of distance make explicit and fundamental use of the point-process nature of spike trains. The formal structure of this approach is an embedding of spike trains into a "metric space." Metric spaces have well-defined distances but do not require the assumption of a linear structure (with implications for superposition and scaling) that would be inherent had we embedded spike trains into a vector space. As pointed out recently by Hopfield (18), there are good theoretical reasons to question whether such vector-space assumptions are appropriate. Indeed, the metrics we consider, which follow naturally from neurobiological heuristics, are *in*consistent with a linear structure. While one can use vector-space embeddings to visualize the metrical relationships between spike trains (Figure 9, Figure 14D - E, Figure 15C - E, Figure 16B - D), the high dimensionality and presence of hyperbolic coordinates in these embeddings indicates that the vector space approximation is a poor one.

This approach is fundamentally hypothesis-driven and in principle can be applied to any proposal for temporal coding, provided that it can be formalized in terms of a metric, or a family of metrics that depends on a parameter (such as our cost $q$). Comparison of clustering across distinct families of metrics provides qualitative insight into the nature of coding, while comparison of clustering within a parametric family provides quantitative insight (such as the "precision" of coding). We cannot prove that this approach will encompass any notion of temporal coding ("encoding" as defined by Theunissen and Miller (44)), but, arguably, the notion of a metric is a *sine qua non* of any rigorous hypothesis for temporal coding. Certainly, this approach is at least as general as are methods that require vector-space embedding, since any such embedding automatically provides a metric (the Euclidean distance).

The metrics we have considered induce distinct topologies on the space of spike trains. Some of the metrics we consider are related by a strict topological hierarchy (Figure 1). Indeed, adding a new kind of step to the definition of a metric always leads to a metric which is coarser than (i.e., refined by) the original metric. However, as the simulations and the analyses of experimental data show, topological refinement does not necessarily imply greater clustering. Moreover, parametric changes in the metric (i.e., variations in $q$) result in no change in the induced topology but frequently in a dramatic change in the degree of clustering.

Bialek's reconstruction method (5) rests on a search for a function $L(t)$ that, when convolved with a spike train, provides an optimal reconstruction of a temporal input. Each candidate function $L(t)$ defines a distance (a Euclidean distance of spike trains convolved with $L(t)$), and the reconstruction approach can be viewed as a minimization of such a distance. Except for "pathological" cases, the distances corresponding to each choice for $L(t)$ have equivalent topologies (which are also equivalent to the one induced by $D^{spike}$). Thus, in both approaches, the characterization of a temporal code appears to require more than merely the topology induced by a metric.

*Variety of behaviors revealed by this approach*

For the two families of metrics we focussed on ($D^{spike}$ and $D^{interval}$), our methods readily reveal a variety of behaviors in simulated datasets. This ranges from no (Figure 4A - D, Figure 6A - D) or minimal (Figure 4E - F) evidence of temporal coding, to dramatic encoding by spike times and intervals (Figure 5E - H), to encoding primarily by spike times and not by intervals (Figure 5A - D), to encoding primarily by intervals and not by spike times (Figure 6E - H, Figure 7). Figure 7 also demonstrates that the present approach, without further machinery, suffices to identify coding through the genesis of deterministic chaos. These qualitative behaviors are largely independent of the exponent $z$ used for clustering.

In all of the analyses, the estimate of chance clustering $H_0$ is relatively insensitive to the parameter of temporal precision, $q$. This underscores the point that the temporal resolution of the analysis is achieved without binning, and hence without an artifactual increase in apparent information despite the increase in embedding dimension at high values of $q$ (Figure 11). This is of particular practical importance given the millisecond precision (23, 24) that cortical neurons possess. The present approach provides a practical way to determine whether spike times, at this level of precision, carry information concerning the stimulus, and thus to differentiate between intrinsic precision and informative precision. The large optimal value of $q$ for auditory data (Figure 13) indicates a role for this high degree of temporal precision in this sensory modality. However, our analysis also indicates that this is not universal: for neurons in visual cortex, maximal clustering is achieved for values of $q$ that correspond to temporal precisions on the order of 20 msec (Figure 14, Figure 15) or more (47), in agreement with measures of temporal precision achieved through more traditional analyses of temporal coding (17).

*Comparison with other clustering methods of analysis of temporal coding*

Middlebrooks et al. (28) used a neural network approach to cluster spike trains and to demonstrate temporal coding of the azimuth of a sound's origin in the responses of single neurons in auditory cortex. Our finding of an increase in $H$ for $q > 0$ (Figure 13A) is in agreement with the authors' inference (28) of temporal coding, but we believe that our approach adds to the neural-network analysis carried out by the authors in several ways. The peak in $H$ near $q \approx 256$ provides an estimate of 3 - 4 msec for the jitter of the temporal structure. Comparison of the difference $H - H_0$ for both $D^{spike}[q]$ and $D^{interval:min}[q]$ shows that the pattern of absolute spike times, but not of spike intervals, is systematically dependent on the stimulus. Finally, multidimensional scaling (Figure 13C and D) of the neural responses according to the distances $D^{spike}[q]$ that are associated with strong clustering provides a vivid demonstration of the representation of the azimuth of a sound's origin in the temporal structure of the neural response.

In comparison with the information analyses of temporal coding considered by Golomb et al. (14), the present approach yields lower values of $H$ and $H - H_0$ by 0.05 - 0.1 (up to 20%). Presumably, this is a consequence of two features of our approach: (i) We used a clustering method which made no assumptions about the shapes of clusters. This has the advantage of identifying clusters even for spike train loci that are concentric (Figure 6, Figure 9), but it would likely provide a less-efficient clustering scheme in situations in which spike train loci conformed to relatively uniform clouds. (ii) The clustering was based on metrics $D^{spike}$ and $D^{interval:min}$, which were chosen to test specific hypotheses concerning the nature of temporal coding, rather than attempting to optimize them to separate spike trains which were created from time-dependent Poisson processes.

Despite these differences, there is a substantial similarity in the biological conclusions that one would draw from the two calculations: the greater information in the parvocellular neuron's responses, and the greater relative importance of temporal coding in the magnocellular neuron's responses. This is reassuring for potential users of either approach. The somewhat smaller value of $H$ and $H - H_0$ that we obtain is a combination of the price paid for a more generic clustering algorithm and the deviation of the metrics $D^{spike}$ and $D^{interval:min}$ from the geometry that underlies the fundamentally linear simulation. That is, the difference in apparent clustering is a window on the underlying geometry of the simulated spike trains themselves. It remains to be seen whether a similar difference is generally found in the analysis of real neuronal discharges.

## ACKNOWLEDGEMENTS

The authors thank Peter Sellers for his insights and helpful discussions. A portion of this work was presented at the 1994 meeting of the Society for Neuroscience, Miami (46). This work was supported by NIH grants EY9314 (JDV) and NS01677 (KPP), The McDonnell-Pew Foundation (KPP), The Revson Foundation (KPP), and The Hirschl Trust (JDV). We thank Mary Conte for her expert technical assistance, Chiye Aoki and Teresa Milner for advice and assistance with histology, and Danny Reich for his comments on the manuscript.

# BIBLIOGRAPHY

1. Abeles, M. (1982) Local Cortical Circuits, an Electrophysiological Study. Springer, Berlin.

2. Abeles, M. (1982) Role of the cortical neuron: integrator or coincidence detector? Isr. J. Med. Sci. 18, 83-92.

3. Abeles, M., & Gerstein, G.L. (1988) Detecting spatiotemporal firing patterns among simultaneously recorded single neurons. J. Neurophysiol. 60, 909-924.

4. Abramson, N. (1963) Information Theory and Coding. McGraw-Hill, New York.

5. Bialek, W., Rieke, F., de Ruyter van Steveninck, R.R., & Warland, D. (1991) Reading a neural code. Science 252, 1854 - 1857.

6. Bliss, T.V.P. & Collingridge, G.L. (1993) A synaptic model of memory: long-term potentiation in the hippocampus. Nature 361, 31-39.

7. Carlton, A.G. (1969) On the bias of information estimates. Psychological Bulletin 71, 108-109.

8. Chee-Orts, M-N., & Optican, L.M. (1993) Cluster method for analysis of transmitted information in multivariate neuronal data. Biol. Cybern. 69, 29-35.

9. Fagen, R.M. (1978) Information measures: statistical confidence limits and inference. J. Theor. Biol. 73, 61-79.

10. Fukunaga, K. (1990) Introduction to Statistical Pattern Recognition, 2nd ed. Academic Press, NY. 591 pp.

11. Gaal, S.A. (1964). Point Set Topology. Academic Press, NY. 317 pp.

12. Gawne, T.J., McClurkin, J.W., Richmond, B.J. , & Optican, L.M. (1991) Lateral geniculate neurons in behaving primates. III. Response predictions of a channel model with multiple spatial-to-temporal filters. J. Neurophysiol. 66, 809-823.

13. Geisler, W.S., Albrecht, D.G., Salvi, R.J., & Saunders, S.S. (1991) Discrimination performance of single

neurons: rate and temporal-pattern information. J. Neurophysiol. 66, 334-362.

14. Golomb, D., Hertz, J., Panzeri, S., Richmond, B., & Treves, A. (1996) How well can we estimate the information carried in neuronal responses from limited samples? Neural Computation, in press.

15. Golomb, D., Kleinfeld, D., Reid, R.C., Shapley, R.M., & Shraiman, B.I. (1994) On temporal codes and the spatiotemporal response of neurons in the lateral geniculate nucleus. J. Neurophysiol. 72, 2990-3003.

16. Green, P.E. (1978) Analyzing Multivariate Data. Dryden Press, Hinsdale, IL. 519 pp.

17. Heller, J., Hertz, J.A., Kjaer, T.W., & Richmond, B.J. (1995) Information flow and temporal coding in primate pattern vision. J. Computational Neuroscience 2, 175-193.

18. Hopfield, J. J. (1995) Pattern recognition computation using action potential timing for stimulus representation. Nature 376, 33 - 36.

19. Kjaer, T.W., Hertz, J.A., & Richmond, B. J. (1994) Decoding cortical neuronal signals: network models, information estimation, and spatial tuning. J. Computational Neuroscience 1, 109-139.

20. Kruskal, J.B., & Wish, M. (1978) Multidimensional Scaling. Sage Publications, Beverly Hills. 93 pp.

21. Larson, J., Wong, D. & Lynch, G. (1986) Patterned stimulation at the theta frequency is optimal for the induction of hippocampal long-term potentiation. Brain Res. 368, 347-350.

22. Laurent, G., Wehr, M., & Davidowitz, H. (1996) Temporal representations of odors in an olfactory network. J. Neurosci. 16, 3837-3847.

23. Lestienne, R. (1996) Determination of the precision of spike timing in the visual cortex of anaesthetised cats. Biol Cybern. 74, 55-61.

24. Mainen, Z., & Sejnowski, T.J. (1995) Reliability of spike timing in neocortical neurons. Science 268, 1503-1506.

25. McClurkin, J.W., Gawne, T.J., Optican, L.M., & Richmond, B.J. (1991) Lateral geniculate neurons in behaving primates. II. Encoding information in the temporal shape of the response. J. Neurophysiol. 66, 794-

808.

26. McClurkin, J.W., Optican, L.M., Richmond, B.J, & Gawne, T.J. (1991) Concurrent processing and complexity of temporally coded messages in visual perception. Science 253, 675-677.

27. Mel, B.W. (1993) Synaptic integration in an excitable dendritic tree. J. Neurophysiol. 70, 1086-1101.

28. Middlebrooks, J.C., Clock, A.E., Xu, L., & Green, D.M. (1994) A panoramic code for sound location by cortical neurons. Science 264, 842-844.

29. Optican, L.M., Gawne, T.J., Richmond, B.J., & Joseph, P.J. (1991) Unbiased measures of transmitted information and channel capacity from multivariate neuronal data. Biol. Cybern. 65, 305-310.

30. Optican, L.M., & Richmond, B.J. (1987) Temporal encoding of two-dimensional patterns by single units in primate inferior temporal cortex. III. Information theoretic analysis. J. Neurophysiol. 57, 162-178.

31. Perkel, D.H., Gerstein, G.L., & Moore, G.P. (1967) Neuronal spike trains and stochastic point processes. I. The single spike train. Biophys. J. 7, 391-418.

32. Perkel, D.H., Gerstein, G.L., & Moore, G.P. (1967) Neuronal spike trains and stochastic point processes. II. Simultaneous spike trains. Biophys. J. 7, 419-440.

33. Purpura, K., Chee-Orts, M.N. & Optican, L.M. (1993) Temporal encoding of texture properties in visual cortex of awake monkey. Society for Neuroscience Abstracts 19, 771.

34. Rapp, P.E., Zimmerman, I.D., Vining, E.P., Cohen, N., Albano, A.M., and Jiménez-Montaño, M. A. (1994) The algorithmic complexity of neural spike trains increases during focal seizures. Journal of Neuroscience 14, 4731-4739.

35. Richmond, B.J., Optican, L.M., Podell, M., & Spitzer, H. (1987) Temporal encoding of two-dimensional patterns by single units in primate inferior temporal cortex. I. Response characteristics. J. Neurophysiol. 57, 132-146.

36. Rose, G.M., & Dunwiddie, T.V. (1986) Induction of hippocampal long-term potentiation using

physiologically patterned stimulation. Neurosci. Lett.69 , 244-248.

37. Sellers, P. H. (1974) On the theory and computation of evolutionary distances. SIAM J. Appl. Math. 26, 787-793.

38. Sellers, P. H. (1979) Combinatorial complexes. Reidel, Dordrecht, Holland, 184 pp.

39. Sen, K., Jorge-Rivera, J.C., Marder, E., & Abbott, L.F. (1996) Decoding synapses. J. Neuroscience 16, 6307-6318.

40. Sherman, S.M., & Koch, C. (1986) The control of retinogeniculate transmission in the mammalian lateral geniculate nucleus. Exp. Brain Res. 63, 1-20.

41. Skarda, C.A., & Freeman, W.J. (1987) How brains make chaos to make sense of the world. Behavioral and Brain Sciences 10, 161-195.

42. Softky, W.R., & Koch, C. (1993) The highly irregular firing of cortical cells is inconsistent with temporal integration of random EPSP's. J. Neurosci. 13, 334-350.

43. Softky, W.R. (1995) Simple codes versus efficient codes. Current Opinion in Neurobiology 5, 239-247.

44. Theunissen, F., & Miller, J. P. (1995) Temporal encoding in nervous systems: a rigorous definition. J. Comput. Neuroscience 2, 149-162.

45. Treves, A., & Panzeri, S. (1995) The upward bias in measures of information derived from limited data samples. Neural Computation 7, 399-407.

46. Victor, J.D., & Purpura, K. (1994) A new approach to the analysis of spike discharges: cost-based metrics. Society for Neuroscience Abstracts 20, 315.

47. Victor, J.D., and Purpura, K. (1996) Nature and precision of temporal coding in visual cortex: a metric-space analysis. J. Neurophysiol. 76, 1310-1326.

48. Victor, J.D., Purpura, K., Katz, E., & Mao, B. (1994) Population encoding of spatial frequency, orientation, and color in macaque V1. J. Neurophysiol. 72, 2151-2166.

## FIGURE AND TABLE LEGENDS

Figure 1. Topological relationships among several families of distances. An arrow leading from one family to another family means that the higher family is a topological refinement of the lower family. The distances are defined in the text.

Figure 2. A minimal-cost path of elementary steps associated with $D^{spike}[q]$ connecting two spike trains $S_a$ and $S_b$. Reproduced from (47).

Figure 3. Examples for which a simple recursive algorithm would fail to calculate the minimal path for distances $D[q_{spike}, q_{interval}]$.

Figure 4. Simulation of rate discrimination. Information-theoretic measure of clustering $H$ derived from $D^{spike}[q]$ (solid lines) and estimates of chance information $H_0$ derived from resampled datasets (broken lines, open symbols), as a function of cost/sec ($q$) and clustering exponent ($z$). Calculations are based on simulated responses to five stimulus classes, which elicited steady firing with mean rates of 2, 4, 6, 8, and 10 impulses/sec. Panels A-D: Poisson process (c.v. = 1.0); clustering exponent $z$ = -8, -2, 2, and 8. Panels E-H: Iterated Poisson process (c.v. = 0.125); clustering exponent $z$ = -8, -2, 2, and 8. Error bars represent ± 2 s.e.m. The maximum possible value of $H$ (arrow labelled "ideal") is $\log_2(5) \approx 2.32$.

Figure 5. Simulation of temporal phase discrimination. Information-theoretic measure of clustering $H$ derived from $D^{spike}[q]$ (filled circles), $D^{interval:fix}[q]$ (filled squares), and $D^{interval:min}[q]$ (filled triangles), as a function of cost/sec ($q$) and clustering exponent ($z$). Estimates of chance information $H_0$ derived from resampled datasets (broken lines, open symbols). Calculations are based on simulated responses to four stimulus classes, which elicited sinusoidally-modulated firing at each of four phases. Panels A-D: Poisson process (c.v. = 1.0); clustering exponent $z$ = -8, -2, 2, and 8. Panels E-H: Iterated Poisson process (c.v. = 0.125); clustering exponent $z$ = -8, -2, 2, and 8. Error bars represent ± 2 s.e.m. The maximum possible value of $H$ (arrow labelled "ideal") is $\log_2(4) = 2$.

Figure 6. Simulation of temporal frequency discrimination. Information-theoretic measure of clustering $H$ derived from $D^{spike}[q]$ (filled circles), $D^{interval:fix}[q]$ (filled squares) and $D^{interval:min}[q]$ (filled triangles), as a function of cost/sec ($q$) and clustering exponent ($z$). Estimates of chance information $H_0$ derived from resampled datasets (broken lines, open symbols). Calculations are based on simulated responses to four stimulus classes, which elicited steady firing and sinusoidally-modulated firing at each of three frequencies (but random phases). Panels A-D: Poisson process (c.v. = 1.0); clustering exponent $z$ = -8, -2, 2, and 8. Panels E-H: Iterated Poisson process (c.v. = 0.125); clustering exponent $z$ = -8, -2, 2, and 8. Error bars represent ± 2 s.e.m. The maximum possible value of $H$ (arrow labelled "ideal") is $\log_2(4) = 2$. The value of $H$ that corresponds to perfect discrimination of responses to modulated and unmodulated stimuli (but confusion among the three modulated stimuli) is labelled "detection of modulation", and is given by $(1/4)\log_2(4) + (3/4)\log_2(4/3) \approx 0.81$.

Figure 7. Simulation of discrimination of random and chaotic firing patterns. Information-theoretic measure of clustering $H$ derived from $D^{spike}[q]$ (filled circles), $D^{interval:fix}[q]$ (filled squares) and $D^{interval:min}[q]$ (filled triangles), as a function of cost/sec ($q$) and clustering exponent ($z$). Estimates of chance information $H_0$ derived from resampled datasets (broken lines, open symbols). Calculations are based on simulated responses to two stimulus classes, one of which contained uniformly-distributed interspike intervals in random

order, and the other of which contained the same distribution of interspike intervals, but was governed by a chaotic recursion rule (eq. (8)). Panels A-D: clustering exponent $z$ = -8, -2, 2, and 8. Error bars represent ± 2 s.e.m. The maximum possible value of $H$ (arrow labelled "ideal") is $\log_2(2) = 1$.

Figure 8. Information-theoretic measure of clustering $H$ derived from $D^{spike}[q]$ (filled circles, solid lines) and estimates of chance information $H_0$ derived from resampled datasets (open circles, broken lines), as a function of cost/sec ($q$), for simulated responses to two stimulus classes, which elicited steady firing with mean rates of 6 and 7 impulses/sec. Iterated Poisson process (c.v. = 0.125); clustering exponent $z$ = -2 . Error bars represent ± 2 s.e.m. The maximum possible value of $H$ (arrow labelled "ideal") is $\log_2(2) = 1$.

Figure 9. Multidimensional scaling of simulated responses to two stimulus classes, which elicited steady firing with mean rates of 6 and 7 impulses/sec (Figure 8). $q$ = 1 (A), $q$ = 16 (B), and $q$ = 256 (C). For each value of $q$, the responses to the two stimulus classes are subjected to multidimensional scaling independently (insets) and jointly (main scattergram).

Figure 10. Dimension index (eq. (10)) for the embeddings of Figure 9. Filled symbols: responses to the stimulus which elicited a mean of 6 impulses/sec analyzed alone; open symbols: responses to the stimulus which elicited a mean of 7 impulses/sec analyzed alone; thick line without symbols: all responses analyzed together.

Figure 11. Dimension index (eq. (10)) for embedding sets of spike trains whose interspike intervals are determined by a chaotic process (eq. (8)) (open symbols), a random process (closed symbols), and the two classes combined (thick line without symbols). Panel A: $D^{spike}[q]$. Panel B: $D^{interval:fix}[q]$. Panel C:

$D^{interval:min}[q]$. As in Figure 7, each class had 100 spike trains.

Figure 12. Clustering of simulated (15) responses of model lateral geniculate neurons to Walsh function patterns. Panels A and B: Information-theoretic measure of clustering $H$ derived from $D^{spike}[q]$ (filled circles) and $D^{interval:min}[q]$ (filled triangles), as a function of cost/sec ($q$). Clustering exponent $z = 2$. Estimates of chance information $H_0$ derived from resampled datasets (broken lines, open symbols). Calculations are based on 64 simulated responses to each of 32 Walsh patterns. Error bars represent ± 2 s.e.m. Panels C and D: Dependence of $H$ (solid lines, small filled symbols), $H_0$ (broken lines, small open symbols) and $H$ - $H_0$ (solid lines, large filled symbols) on the number of responses per stimulus class. $D^{count}$ (asterisks), $D^{spike}[q]$ (circles), and $D^{interval:min}[q]$ (triangles). For $D^{spike}[q]$ and $D^{interval:min}[q]$, $q$ is chosen to yield the maximum value of $H$. The maximum possible value of $H$ is $\log_2(32) = 5$.

Figure 13. Analysis of temporal coding of azimuth in cat anterior ectosylvian cortex (data of Middlebrooks et al. (28)). 30 responses to each of 6 ranges of azimuth are analyzed. Panel A: information-theoretic measure of clustering, $H$ (solid lines) for $D^{spike}[q]$ (filled circles) and $D^{interval:min}[q]$ (filled triangles), and estimates of chance information $H_0$ (broken lines, open symbols, error bars represent mean ± 2 s.e.m.) derived from 10 resampled datasets. Panel B: eigenvalues of the matrix of eq. (9) for $D^{spike}[256]$. Panel C: multidimensional scaling for $D^{spike}[256]$, with each spike train's locus projected onto the plane of the first two eigenvectors of the matrix of eq. (9). Each point corresponds to an individual response, and the points are color-coded to indicate the azimuth of the stimulus that elicited the corresponding response. Panel D: the centers of each of the response clouds in Panel C.

Figure 14. Analysis of temporal coding of spatial frequency and orientation in macaque primary visual

cortex. 16 responses to each of 40 stimuli (five spatial frequencies, eight orientations) are analyzed. Panel A: information-theoretic measure of clustering, $H$ (solid lines) for $D^{spike}[q]$ (filled circles) and $D^{interval:min}[q]$ (filled triangles), and estimates of chance information $H_0$ (broken lines, open symbols, error bars represent mean ± 2 s.e.m.) derived from 10 resampled datasets. Panel B: eigenvalues of the matrix of eq. (9) for $D^{spike}[32]$. Panels C, D, and E: multidimensional scaling for $D^{spike}[32]$, with each spike train's locus projected onto the planes determined by pairs of the first three eigenvectors of the matrix of eq. (9). Each point represents the mean position of 16 responses. Trajectories are color-coded to correspond to the five spatial frequencies used in the experiment, and for each spatial frequency, the response at the preferred orientation is marked with a black symbol. Recording H30011.

Figure 15. Analysis of temporal coding of contrast and spatial frequency in macaque primary visual cortex. 128 responses to each of 15 stimuli (three spatial frequencies, five contrasts) are analyzed. Panel A: information-theoretic measure of clustering, $H$ (solid lines) for $D^{spike}[q]$ (filled circles) and $D^{interval:min}[q]$ (filled triangles), and estimates of chance information $H_0$ (broken lines, open symbols, error bars represent mean ± 2 s.e.m.) derived from 10 resampled datasets. Panel B: eigenvalues of the matrix of eq. (9) for $D^{spike}[64]$. Panels C, D, and E: multidimensional scaling for $D^{spike}[64]$. Each spike train's locus is projected onto the planes determined by pairs of the first three eigenvectors of the matrix of eq. (9), and the mean position of 128 responses to each stimulus is plotted. Trajectories are color-coded to correspond to the three spatial frequencies used in the experiment, and for each spatial frequency, the response at the highest contrast is marked with a black symbol. Recording 19/12.

Figure 16. Detailed analysis of the data of of Figure 15 for $D^{interval:min}[64]$. Panel A: eigenvalues of the matrix of eq. (9). Panels B, C, and D: multidimensional scaling. Each spike train's locus is projected onto the planes determined by pairs of the first three eigenvalues, and the mean position of 128 responses to each stimulus is plotted. As in Figure 15, trajectories are color-coded to correspond to the three spatial frequencies used in the experiment, and for each spatial frequency, the response at the highest contrast is marked with a black symbol. Recording 19/12.